\documentclass[aps,prl,showpacs,twocolumn,groupedaddress,floatfix]{revtex4}
\usepackage{amsmath}
\usepackage{graphicx}
\usepackage{color}

\def\beq{\begin{eqnarray}}
\def\eeq{\end{eqnarray}}

\begin{document}

\title{A variational sinc collocation method for strong-coupling problems}
\author{Paolo Amore}\email{paolo@ucol.mx} 
\affiliation{Facultad de Ciencias, Universidad de Colima, \\
Bernal D\'{\i}az del Castillo 340, Colima, Colima, Mexico} 

\date{\today}

\begin{abstract}    
We have devised a variational sinc collocation method (VSCM) which can be used to 
obtain accurate numerical solutions to many strong-coupling problems. Sinc functions 
with an optimal grid spacing are used to solve the linear and non-linear Schr\"odinger 
equations and a lattice $\phi^4$ model in $(1+1)$. 
Our results indicate that errors decrease exponentially with the number of grid 
points and that a limited numerical effort is needed to reach high precision.
\end{abstract}
\maketitle


Due to the inapplicability of perturbation theory in the strong coupling regime, a 
number of different techniques has been devised in the past to deal with strong--coupling 
problems. A particular attention has gone into developing new methods, in which 
variational principles are used to improve perturbation theory, leading to results 
which are valid on a much larger domain. The Linear Delta Expansion (LDE)\cite{lde} 
and the Variational Perturbation Theory (VPT)\cite{Klei04} are probably the best 
known examples of such efforts: in many cases these methods allow to obtain series 
with finite (or even infinite) radius of convergence, in contrast with the divergent 
series which are usually obtained using perturbation theory\cite{Klei95}. 

In this letter we wish to show that the variational ideas which have inspired both the LDE 
and the VPT methods can be also applied to improve the performance of numerical techniques.
The numerical method that we are using is the Sinc Collocation (SCM)\cite{Steng93}, which 
uses sinc functions to efficiently ``discretize'' a problem in given region. Sinc functions 
are used in different areas of physics and  mathematics ( see for example \cite{sinc} and 
references therein).

A sinc function is defined as 
\beq
S_k(h,x) &\equiv& \frac{\sin\left(\pi (x-k h)/h \right)}{\pi (x-k h)/h}  \ .
\label{eq_1_1}
\eeq
and obeys the integral representation
\beq
S_k(h,x) &=& \frac{h}{2\pi} \int_{-\pi/h}^{+\pi/h} \ e^{\pm i (x- k h) t} \ dt \ .
\label{eq_1_2}
\eeq

The reader interested in a more detailed account of the properties of the 
sinc function should refer to \cite{Steng93}; here we only state the main 
properties which will turn useful in the following.

Using eq.~(\ref{eq_1_2}) it is straightforward to evaluate the integrals
\beq
\label{eq_1_3}
{\cal I}_1 &\equiv& \int_{-\infty}^{+\infty} S_k(h,x) \ dx = h  \\
{\cal I}_2 &\equiv& \int_{-\infty}^{+\infty} S_k(h,x) \ S_l(h,x) \ dx = h \ \delta_{kl}  .
\eeq

A function $f(x)$ analytic on a rectangular strip centered on the real axis 
can be approximated in terms of sinc functions as
\beq
f(x) \approx  \sum_{k=-\infty}^{+\infty} \ f(k h) \ S_{k}(h,x) \ .
\label{eq_1_4}
\eeq
Using eq.~(\ref{eq_1_4}) together with eq.~(\ref{eq_1_3}) one obtains
\beq
\int_{-\infty}^{+\infty} f(x) \ dx &\approx& h \ \sum_{k=-\infty}^{\infty} f(k,h)  .
\label{eq_1_5}
\eeq

An expression for the error in eq.~(\ref{eq_1_4}) has been obtained by Stenger~\cite{Steng93}, 
showing that it decays exponentially as the spacing $h$ is reduced.

\begin{figure}
\begin{center}
\includegraphics[width=7cm]{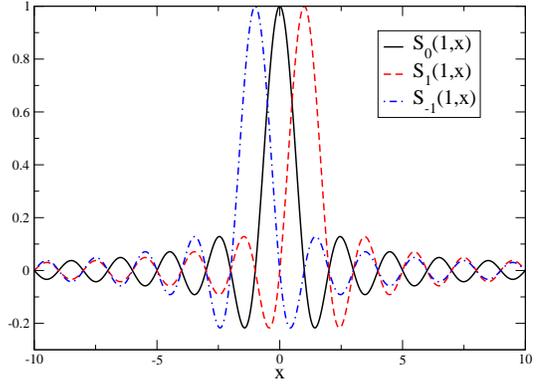}
\caption{Sinc functions corresponding to different values (color online).}
\label{FIG1}
\end{center}
\end{figure}

As the reader can appreciate in Fig.\ref{FIG1} for a fixed $h$ a given sinc function selects 
a point on the real line, corresponding to its maximum value, and vanishes in the points of 
intersection with the other sinc functions. This property, which allows to obtain a proper ``discretization'' of 
a problem in the continuum, is at the basis of the SCM.

We now describe the SCM in detail by considering the stationary Schr\"odinger equation
\beq
- \frac{1}{2} \frac{d^2\psi}{dx^2} + V(x) \psi(x) = E \psi(x) \ .
\eeq

The matrix elements  $H_{kl}$ of the hamiltonian evaluated 
in the set of sinc functions are given by
\beq
H_{kl} \approx \left[-  \frac{1}{2} \ c_{kl}^{(2)} + \delta_{kl} \ V(k h)  \right] \ .
\eeq

Notice that the kinetic term has been obtained by using the property
\beq
\frac{d^2}{dx^2} S_k(h,x)  = \sum_{l=-\infty}^{\infty} \ c_{lk}^{(2)} \  S_k(h,x)  \ ,
\eeq
where
\beq
c_{lk}^{(2)} &=& \left\{ \begin{array}{c}
- \frac{\pi^2}{3 h^2} \ \ if \ \ k=l \\
- \frac{2}{h^2} \frac{(-1)^{k-l}}{(k-l)^2} \ \ if \ \ k \neq l 
\end{array}
\right. \ ,
\eeq
while the potential matrix has been approximated by the diagonal matrix of the 
potential evaluated over the grid.

Once $h$ is specified the diagonalization of $H_{kl}$ allows to obtain 
numerical approximations to the energies and wave functions of the problem: this
strategy was used in \cite{Wei00} to solve the Schr\"odinger equation corresponding
to different potentials. Although the choice of $h$ strongly affects the precision 
of the numerical results, no procedure to determine $h$ is discussed in \cite{Wei00}. 
Using a different method, the author and collaborators~\cite{Am_1} have solved numerically 
the Schr\"odinger equation for the anharmonic oscillator using an arbitrary basis 
of Gauss-Hermite functions, depending upon a scale factor. 
In that paper it was proved that the arbitrary scale factor can be chosen optimally by 
applying the principle of minimal sensitivity (PMS)\cite{Ste81} to the sub-trace of the 
hamiltonian matrix.

\begin{figure}
\begin{center}
\includegraphics[width=7cm]{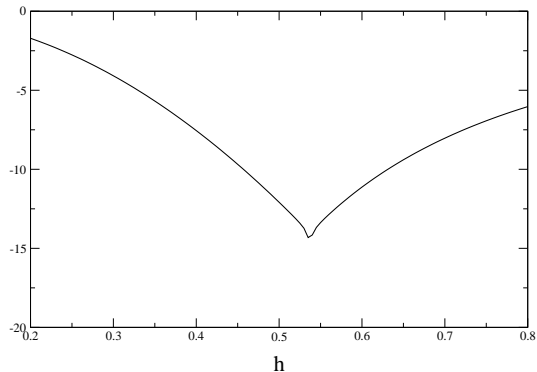}
\caption{$\log_{10}|E_0-E_0^{exact}|$ as a function of the spacing $h$ using $k_{max}=10$ for the harmonic oscillator
$V(x) = x^2/2$.}
\label{FIG2a}
\end{center}
\end{figure}

Using the same procedure we regard $h$ as a variational parameter and consider the trace
\beq
Tr\left[H\right] = \frac{\pi^2}{6 h^2} \ (2 k_{max}+1) + \sum_{k=-k_{max}}^{k_{max}} V(k h) \ , 
\eeq
where $2 k_{max}+1$ is the number of sinc functions (grid points) used in the evaluation.

The solution to the PMS equation
\beq
\frac{d}{dh} Tr\left[H\right] = 0 
\label{pms}
\eeq
provides the optimal spacing~\footnote{Usually this equation admits a unique real solution.}. 
Once that $h$ has been determined using the PMS our method allows to obtain quite rapidly the 
numerical approximations\footnote{All the examples considered in this letter have been obtained using a Mathematica code, running on a Linux desktop, 
with times of execution ranging from few seconds to few minutes.}.

In fig.~\ref{FIG2a} we display the $\log_{10}|E_0-E_0^{exact}|$ as a function of the spacing $h$ using $21$ grid points 
($k_{max}=10$) for the harmonic oscillator $V(x) = x^2/2$. The PMS condition, eq.~(\ref{pms}), yields $h_{PMS} = 0.547$, 
which is remarkably close to the minimum of the curve.

The anharmonic oscillator
\beq
H =  - \frac{d^2}{dx^2} + x^2 +  g \ x^4 \nonumber
\eeq
provides a more demanding test of our method. 
We have obtained the ground state energy corresponding to $g =2000$ 
using a grid of $101$ points ($k_{max} = 50$) and compared it with 
the precise results of \cite{mei97}, finding that the first $42$ digits
are correct (underlined):
\begin{widetext}
\beq
E_0 &=& \underline{13.3884417010080619390061769028072865229609}9 \nonumber \ .
\eeq
\end{widetext}
The result is also seen to converge exponentially to the exact answer as a 
function of the number of grid points.

We now consider the Gross-Pitaevskii (GP) equation
\beq
\left[ - \frac{1}{2} \ \frac{d^2}{dx^2} + V_{ext}(x) + 4 \ \pi  \ a \ |\psi(x)|^2 \right] \psi(x) = E \ \psi(x) \nonumber \ ,
\eeq
which is relevant in the study of Bose-Einstein condensation. $V_{ext}(x)$ is the external confining potential and $E$ is the energy
of the condensate. The wave function $\psi(x)$ is normalized to yield the number of particles in the condensate, i.e.
$\int dx \ |\psi(x)|^2 = N$.

We have applied our method to the GP equation by first solving the corresponding linear equation, and by 
then implementing a self-consistent procedure in which the density term is evaluated taking the wave function 
calculated at the previous step and then it is used to build an effective potential 
$V_{eff}(x) \equiv V_{ext}(x)  +  4 \ \pi  \ a \ |\psi(x)|^2$.
In this potential  the resulting Schr\"odinger equation is solved again and the procedure is iterated until 
self--consistency is reached. 

\begin{figure}
\begin{center}
\includegraphics[width=7cm]{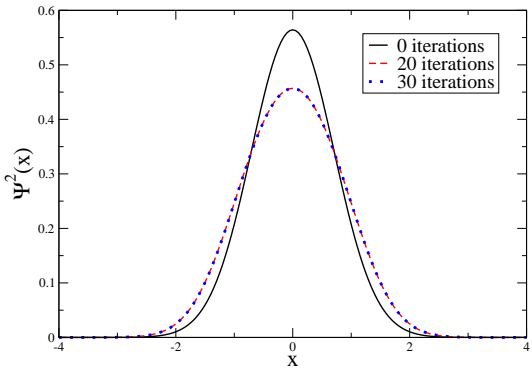}
\caption{Probability density for the  ground state of the GP equation using $V_{ext}(x)= x^2/2$ (color online).}
\label{FIG3}
\end{center}
\end{figure}

\begin{figure}
\begin{center}
\includegraphics[width=7cm]{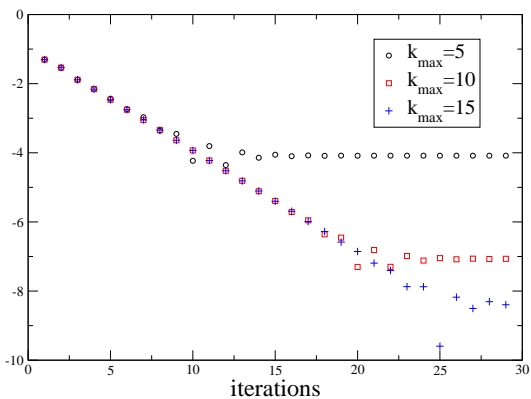}
\caption{$\log_{10} |E_0 -E_0^{exact}|$ as a function of the number of iterations for different grid sizes (color online).}
\label{FIG4}
\end{center}
\end{figure}

In fig.~\ref{FIG3} we have plotted the wave function in an harmonic trap, obtained after $0$, $20$ and $30$ iterations of our method, 
assuming $4\pi a=1$ and $\int_{-\infty}^{+\infty} |\psi(x)|^2 \ dx = 2$. A grid of $21$ points has been used.
The $0$th order wave function corresponds to the usual harmonic oscillator wave function. 
In fig.~\ref{FIG4} we have plotted $\log_{10} |E_0 -E_0^{exact}|$ as a function of the number of iterations for different grid sizes
($11$, $21$ and $31$ respectively): the datas initially display an exponential decay -- independent of the grid size -- 
which is then followed by a plateau. The plateau signals that the maximal precision has been achieved for a given grid size.

As a last example of application of our method we consider a lattice $\phi^4$ in $1+1$ dimensions. This model 
has been studied by Nishiyama in ~\cite{Nish01} and corresponds to the hamiltonian
\beq
{\cal H} = \sum_i \left[ \frac{\pi_i^2}{2} + \frac{1}{2} \ \left(\phi_i-\phi_{i+1}\right)^2 +\frac{1}{2} \phi_i^2 + g \phi_i^4 \right] \ .
\eeq
The fields obey the canonical commutation relations $\left[\phi_i, \pi_j \right] = i \delta_{ij}$ 
and $\left[\phi_i, \phi_j \right] = \left[\pi_i, \pi_j \right] = 0$. 
Following \cite{Nish01} we perform a rescaling of the fields $\phi \rightarrow g^{-1/6} \phi$ and $\pi \rightarrow g^{1/6} \pi$  
and obtain
\beq
{\cal H} = g^{1/3} \ 
\sum_i \ \left[ \frac{\pi_i^2}{2} + \phi_i^4 + \lambda \ \left( \frac{1}{2} \left(\phi_i-\phi_{i+1}\right)^2 + \frac{\phi_i^2}{2}\right) 
\right] \nonumber \ ,
\eeq
where $\lambda \equiv g^{-2/3}$. We express the ground state energy as $E_g = g^{1/3} \ \epsilon_g$.

Nishiyama has numerically solved this model using a Linked Cluster Expansion (LCE) and the Density Matrix Renormalization
Group (DMRG)\cite{White92}: using the LCE he has obtained a perturbation series in $\lambda$, up to order $11$, whose convergence
has then been improved using Aitken's $\delta^2$ process. The comparison with the DMRG results shows that LCE is 
valid up to $\lambda \approx 2$.

We wish to show that the same problem can be solved using our method. We have proceeded as follows:
first we have solved the Schr\"odinger equation for the anharmonic oscillator, corresponding to setting $\lambda=0$, and we have
obtained the wave function $\Psi(\phi) \approx \sum_{r=-k_{max}}^{k_{max}} \alpha_r \ S_r(h,\phi)$; we have then used 
$\Psi(\phi)$ to evaluate the matrix element $\langle \Psi(\phi_{i+1}) | H | \Psi(\phi_{i+1})\rangle$, obtaining the effective 
potential felt by the $i$th site:
\begin{widetext}
\beq
\tilde{V}(\phi) = \phi^4 + \lambda \ \left[ \frac{1}{2} \left(\phi^2 -  2 \phi \sum_{r=-k_{max}}^{k_{max}} \alpha_r^2 rh 
+ \sum_{r=-k_{max}}^{k_{max}} \alpha_r^2 (rh)^2\right) + \frac{\phi^2}{2} \right] \ .
\eeq
\end{widetext}

The Schr\"odinger equation is then solved again and the coefficients $\alpha_r$ are recalculated. 
The procedure is repeated until self--consistency is reached. 

\begin{figure}
\begin{center}
\includegraphics[width=8cm]{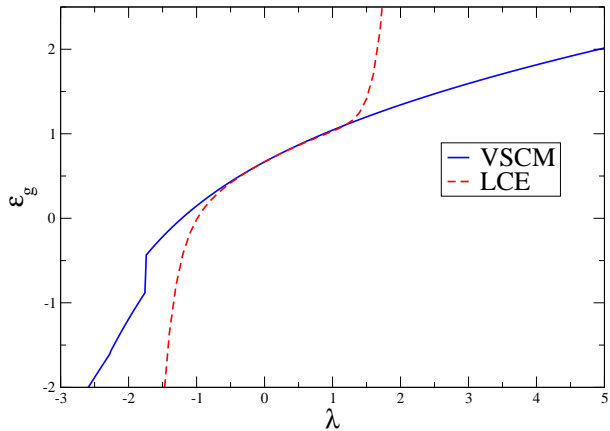}
\caption{Ground state energy for the lattice $\phi^4$ model as a function of $\lambda = g^{-2/3}$ (color online).}
\label{FIG5}
\end{center}
\end{figure}

In fig.~\ref{FIG5} we have calculated the scaled ground state energy, $\epsilon_g$, as a function of $\lambda$,  
using a grid of $41$ grid points ($k_{max}=20$). The solid curve corresponds to the perturbative expansion of 
eq.(6) of \cite{Nish01}, to order $\lambda^{11}$, obtained with the LCE. Our result compares  quite favorably also with 
the results obtained with the DMRG -- see Fig. 2 of \cite{Nish01} -- and have been
obtained in few minutes of running time on a Linux desktop running Mathematica. 

\begin{figure}
\begin{center}
\includegraphics[width=7cm]{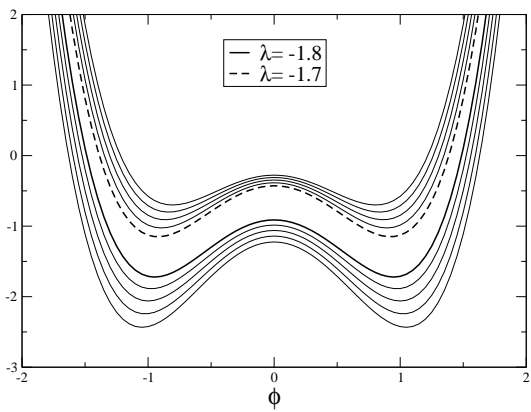}
\caption{Effective potential $\tilde{V}(\phi)$ for values of $\lambda$ close to the discontinuity. The curves 
differ by $\Delta\lambda = 0.1$.}
\label{FIG6}
\end{center}
\end{figure}

Notice that in \cite{Nish01} it was speculated the presence of a singularity around $\lambda=-2$, which is possibly related to 
the onset of a phase transition. Our simulation shows the presence of a discontinuity located at $\lambda \approx -1.75$.
As it can be seen from fig.~\ref{FIG6}, such discontinuity is due to a sudden jump in the effective potential.

We wish to conclude this letter stressing few points, which we believe to be important:  the VSCM 
provides errors which decrease exponentially with the grid size; the grid spacings are obtained using 
the PMS and allow to achieve optimal results with limited numerical effort; the diagonalization
of the hamiltonian matrix provides approximations for the energies and wave functions of part of the spectrum 
and can used to study the time evolution of a wave packet (similarly to what done in \cite{Am_1}); finally,
our method can also be applied {\sl without modifications} to models with non--polynomial interactions. 

The last example shows that VSCM can be a useful tool in the numerical solution of lattice quantum models 
and it could possibly provide an alternative to numerical methods already present on the ``market''.
Future work in this direction is expected.

\begin{acknowledgments}
P.A. acknowledges support of Conacyt grant no. C01-40633/A-1.  

\end{acknowledgments}

\end{document}